\documentclass[twocolumn,aps,prl,reprint,superscriptaddress,citeautoscript]{revtex4-1}

\usepackage{graphicx}
\usepackage{multirow}
\usepackage{makecell}
\usepackage{hyperref}
\usepackage{xcolor}
\setcitestyle{super}
\usepackage{tocloft}
\usepackage{hhline}
\hypersetup{
     colorlinks   = true,
     citecolor    = blue,
     urlcolor     =black
}

\begin{document}

\title{Frequency-dependent \textit{ab initio} Resonance Raman Spectroscopy}

\author{G. Kukucska}
\affiliation{Department of Biological Physics, ELTE E{\" o}tv{\" o}s Lor{\' a}nd University, P{\'a}zm{\'a}ny P{\'e}ter s{\'e}t{\'a}ny 1/A, 1117 Budapest, Hungary }
\author{V. Z{\'o}lyomi}
\affiliation{National Graphene Institute, University of Manchester, Oxford Road, Manchester M13 9PL, UK.}
\author{J. Koltai}
\affiliation{Department of Biological Physics, ELTE E{\" o}tv{\" o}s Lor{\' a}nd University, P{\'a}zm{\'a}ny P{\'e}ter s{\'e}t{\'a}ny 1/A, 1117 Budapest, Hungary }

\date{\today}


\begin{abstract}
We present a new method to compute resonance Raman spectra based on \textit{ab initio} level calculations using the frequency-dependent Placzek approximation. We illustrate the efficiency of our hybrid quantum-classical method by calculating the Raman spectra of several materials with different crystal structures. Results obtained from our approach agree very well with experimental data in the literature. We argue that our method offers an affordable and far more accurate alternative to the widely used static Placzek approximation. 
\end{abstract}

\maketitle

Raman spectroscopy is one of the most versatile non-destructive characterization methods for molecules and solid state systems.\cite{raman_new_1928} The Raman effect allows one to gauge the structural properties of materials through the frequencies of vibrations which can be determined by the difference between the incoming and outgoing photon energy. In addition, resonance effects in the Raman spectra can reveal details of the electronic structure and optical properties of the examined material.\cite{carvalho_symmetry-dependent_2015}  The distribution of spectral weights between the different peaks in the Raman spectra can also carry information about perturbations in the material such as strain or doping\cite{kukucska_theoretical_2017}, or even lattice defects\cite{cancado_quantifying_2011}.
 
Theoretical modeling of Raman spectra is an exceptionally challenging task. Resonant processes are usually described at the semiempirical level, e.g. using the tight-binding model\cite{venezuela_theory_2011,kukucska_resonance_2019}. In contrast, when \textit{ab initio} methods are employed, the calculations are limited to the static approximation where the matter-light interaction is approximated with the response to a static external electric field within\cite{porezag_infrared_1996,lazzeri_first-principles_2003} the static Placzek approximation.\cite{placzek1,placzek2} Raman peak intensities predicted by the static Placzek approximation are fairly accurate for wide gap semiconductors, i.e. when the laser excitation energy is small compared to the optical gap. However, when the gap is comparable to or smaller than the laser excitation energy, it cannot produce accurate relative intensities any more. Moreover, Raman spectra of metallic or semi-metallic systems cannot be calculated in this way, since the response to a static external electric field in defect-free metals is divergent. 

If resonance effects are taken into account in the calculation of the Raman spectrum, theory can make accurate predictions regardless of the electronic properties.\cite{kukucska_theoretical_2017,kukucska_characterization_2018} Furthermore, by calculating laser energy dependent Raman intensities, resonance effects can be studied in the excitation profile. However, available commercial \textit{ab initio} codes only offer to calculate Raman intensities based on the static Placzek approximation, which limits the extent to which Raman intensities can be predicted for the reasons discussed above. 

A few recent works employed many-body theory to compute the frequency-dependent Raman spectrum \cite{miranda_quantum_2017,gillet_ab_2017,wang_strong_2018} taking excitonic effects into account through the Bethe-Salpeter equation (BSE).\cite{salpeter_relativistic_1951} While these methods are able to provide very good accuracy for theoretical predictions, they are limited to small systems due to the extremely high computational demand of many-body calculations.

\begin{figure}
\begin{center}
\includegraphics[scale=.375]{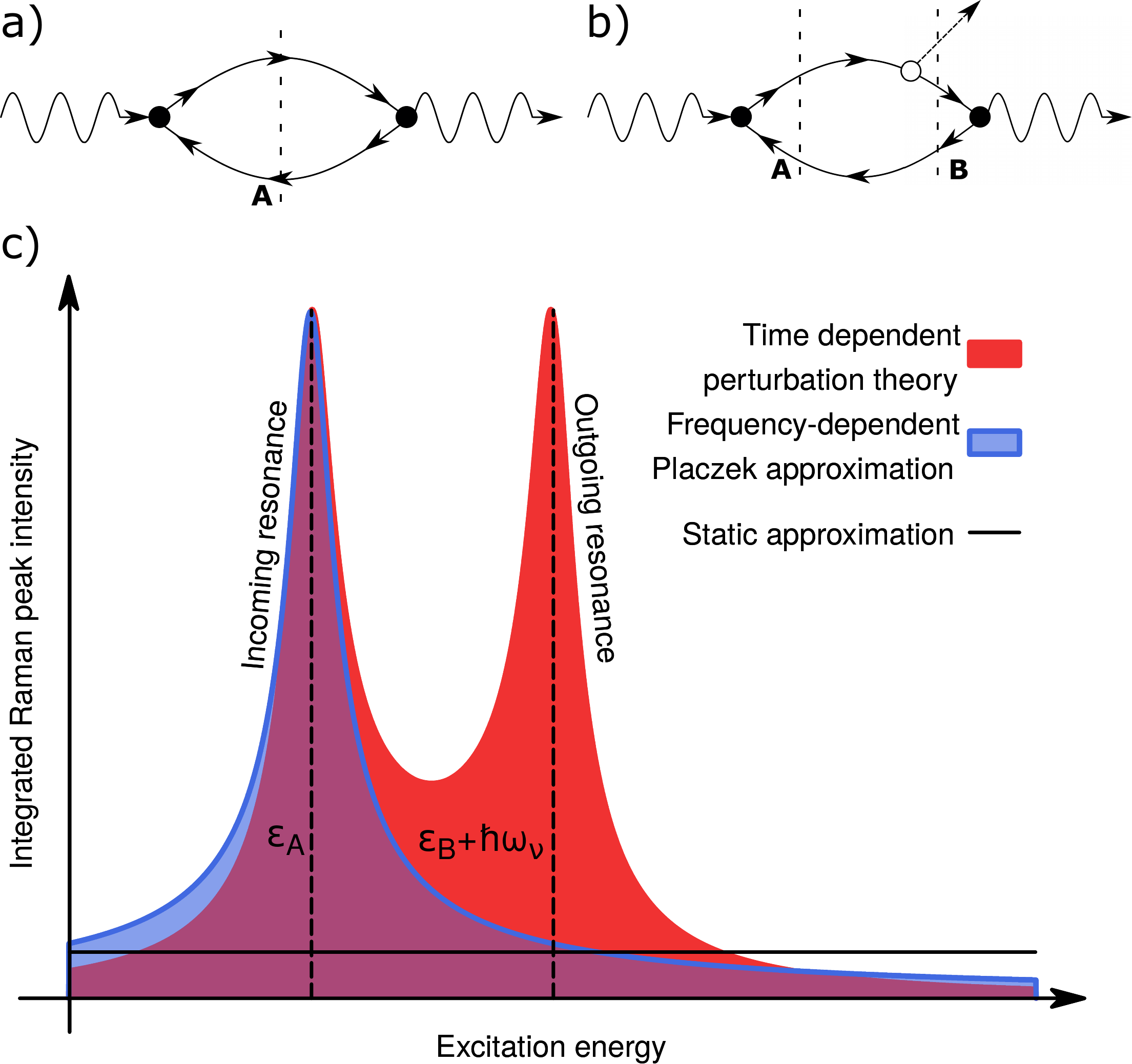}\\
\caption{Feynman diagram of elastic light scattering (a) and the Stokes Raman process (b). Solid lines indicate electron and hole propagation, vertical dashed lines with capital letters denote intermediate virtual states, while wavy solid and straight dotted lines indicate photon and phonon propagation, respectively. Solid and empty vertices denote electron photon and electron phonon interaction, respectively. In (c) we show a schematic representation of how the laser energy dependence of the Raman peak intensity changes when comparing time dependent perturbation theory with the frequency-dependent Placzek approximation and the static approximation (c).\label{compare}}
\end{center}
\end{figure}

In this work we demonstrate a method that enables the efficient computation of resonance Raman spectra at the \textit{ab initio} level using existing computational software. Specifically, we demonstrate how to combine the classical frequency-dependent Placzek approximation with \textit{ab initio} level calculations as implemented in the Vienna \textit{ab initio} Simulation Package (\textsc{VASP}).\cite{kresse_efficient_1996,kresse_ultrasoft_1999} To validate our method we compare our results to Raman spectra calculated using the static approximation implemented in Quantum Espresso (QE)\cite{giannozzi_quantum_2009} and experimental data.\cite{balachandran_raman_1982,gillet_high-temperature_1990,harima_raman_2006,li_bulk_2012,sander_raman_2012,lu_plasma-assisted_2014,hans_agnr_2019} As we will show, our Frequency-dependent \textit{ab initio} Placzek approximation is applicable to any material and provides a fairly accurate description of Raman intensities, without the expense of many-body theory.

To begin, we briefly introduce the theoretical basics of our method. The frequency-dependent Placzek approximation is based on the linear response to a monochromatic electromagnetic field within the long wavelength limit (i.e. neglecting the spatial dependence of the electric field):
\begin{equation}
\mathbf{P}=\hat{\alpha}(\omega_l)\mathbf{E}_0\cos(\omega_l t),
\label{linresp}
\end{equation}
where $\mathbf{E}_0$ is the polarization vector of the incident electric field, $\mathbf{P}$ is the induced polarization and $\hat{\alpha}(\omega_l)$ is the so called polarizability tensor at the laser energy $\omega_l$. In the classical description, Raman intensity is proportional to the derivative of the polarizability with respect to phonon normal modes, resulting in the following expression for the Stokes branch of the Raman spectrum: 
\begin{equation}
I(\omega,\omega_l)=\sum\limits_{\nu}\frac{\omega_s^4}{\omega_{\nu}}\left|\frac{\partial\hat{\alpha}(\omega_l)}{\partial Q_{\nu}}\right|^2\delta(\omega-\omega_{\nu})(n(\omega_\nu)+1),
\label{raman}
\end{equation}
where $\omega=\omega_l-\omega_s$ is the Raman shift, $Q_\nu,\,\omega_\nu$ are the phonon normal mode and frequency, $\omega_s$ is the frequency of the scattered light, $\delta(x)$ is a normalized Lorentzian function and $n(\omega_\nu)$ is the Bose-Einstein distribution at room temperature.\\
In the following we compare the laser energy dependence of amplitudes calculated in this way, with those calculated from time-dependent perturbation theory. Let us consider the expression of the polarizability corresponding to the Feynman diagram of elastic light scattering depicted in Fig. \ref{compare}a:
\begin{equation}
\hat{\alpha}(\omega_l)=\sum\limits_{A}\frac{\langle i|\hat{H}_{e-p}|A\rangle\langle A|\hat{H}_{e-p}|f\rangle}{\hbar\omega_l-\epsilon_A-i\gamma_{e-p}},
\label{eq1}
\end{equation}
where the sum goes over all intermediate virtual state with energy $\epsilon_A$ and wavefunction $|A\rangle$, $\hat{H}_{e-p}$ is the electron-photon interaction and $\gamma_{e-p}$ is the electronic linewidth. The initial $|i\rangle$ and final $|f\rangle$ states are considered to be the electronic and vibrational ground state with zero energy.

To obtain the formal expression of Raman intensities we need to calculate the derivative of this quantity with respect to the normal modes. By displacing atoms according to normal modes, one can see that energies are only perturbed in the second order, whilst the wavefunctions are already perturbed in the first order. The perturbed state $|A'\rangle$ can be expressed as 
\begin{equation}
|A'\rangle=|A\rangle+\sum\limits_{B}\frac{\left\langle B\left|\frac{\partial V_{e-I}}{\partial Q_\nu} Q_\nu\right| A\right\rangle}{\epsilon_A-\epsilon_B-i\gamma_{e-ph}}|B\rangle,
\label{eq2}
\end{equation}

\begin{table*}
\begin{tabular}{|c|c|c|c|c|c|c|c|c|}
\hline
\multirow{2}{*}{ }& \multicolumn{3}{c|}{Frequency}& \multicolumn{3}{c|}{Raman intensity}& \multirow{2}{*}{Symmetry}\\
\cline{2-7}
& \thead{Frequency-dependent\\ \textit{ab initio} Placzek}& \thead{Static\\ Placzek} & Experiment & \thead{Frequency-dependent\\ \textit{ab initio} Placzek}&  \thead{Static\\ Placzek} & Experiment &\\
\hline

\multirow{2}{*}{SiC (fcc)}
 & 970 & 977& 960\cite{harima_raman_2006} & 0.50& 0.50& 0.48\cite{harima_raman_2006} &T$_2$ (LO)\\
\cline{2-8}
 & 797 & 801& 784\cite{harima_raman_2006} & 1.00 & 1.00 & 1.00\cite{harima_raman_2006} &T$_2$ (TO)\\

\hhline{|=|=|=|=|=|=|=|=|}
\multirow{3}{*}{ZnO (hcp)} 
 & 448 & 444& 439\cite{sander_raman_2012} &1.00 & 0.51 &1.00\cite{sander_raman_2012} &E$_2$\\
\cline{2-8}
 & 415 & 414& 412\cite{sander_raman_2012} & 0.04& 0.02 &0.09\cite{sander_raman_2012} &E$_1$\\
\cline{2-8}
 & 397 & 390& 379\cite{sander_raman_2012} & 0.37& 1.00 &0.24\cite{sander_raman_2012} &A$_1$\\

\hhline{|=|=|=|=|=|=|=|=|}
\multirow{4}{*}{SiO$_2$(quartz)}
 & 455 &	458 & 465\cite{gillet_high-temperature_1990} & 1.00 &	1.00 & 1.00\cite{gillet_high-temperature_1990} &	A$_1$\\
\cline{2-8}
 & 255 &	259 & 263\cite{gillet_high-temperature_1990} & 0.04 &	0.02 & 0.11\cite{gillet_high-temperature_1990} &	E\\
\cline{2-8}
 & 225 &	222 & 207\cite{gillet_high-temperature_1990} & 0.19 &	0.20 & 0.20\cite{gillet_high-temperature_1990} &	A$_1$\\
\cline{2-8}
 & 128 &	131 & 129\cite{gillet_high-temperature_1990} & 0.07 &	0.00 & 0.26\cite{gillet_high-temperature_1990} &	E\\

\hhline{|=|=|=|=|=|=|=|=|}
\multirow{5}{*}{Anatase (bcc)}
 & 646 & 680& 640\cite{balachandran_raman_1982} &1.00 & 1.00 & 1.00\cite{balachandran_raman_1982}  & E$_g$\\
\cline{2-8}
 & 505 & 518& 515\cite{balachandran_raman_1982} &0.26 & 0.25 & 0.30\cite{balachandran_raman_1982} &A$_{1g}$\\
\cline{2-8}
 & 503 & 516& 515\cite{balachandran_raman_1982} &0.14 & 0.25 & 0.30\cite{balachandran_raman_1982} &B$_{1g}$\\
\cline{2-8}
 & 371 & 373& 396\cite{balachandran_raman_1982}  &0.72 & 0.09 & 0.66\cite{balachandran_raman_1982}  &B$_{1g}$\\
\cline{2-8}
 & 152 & 137& 147\cite{balachandran_raman_1982} &0.96 & 0.08 & 0.96\cite{balachandran_raman_1982} &E$_{g}$\\

\hhline{|=|=|=|=|=|=|=|=|}
\multirow{2}{*}{Monolayer MoS$_2$}
 & 400 & 407 & 406\cite{li_bulk_2012} & 1.00 & 1.00 & 1.00\cite{li_bulk_2012} &A$_1'$\\
\cline{2-8}
 & 374 & 389 & 382\cite{li_bulk_2012} & 0.92 & 0.43 & 0.94\cite{li_bulk_2012} &E$'$\\

\hhline{|=|=|=|=|=|=|=|=|}
\multirow{3}{*}{Black Phosphorene}
 & 453 & 453& 471\cite{lu_plasma-assisted_2014} & 1.00 & 1.00 & 1.00\cite{lu_plasma-assisted_2014} &A$_{g}$\\
\cline{2-8}
 & 434 & 433& 440\cite{lu_plasma-assisted_2014} & 0.48 & 0.00 & 0.33\cite{lu_plasma-assisted_2014} &B$_{g}$\\
\cline{2-8}
 & 364 & 364& 363\cite{lu_plasma-assisted_2014} & 0.02 & 0.59 & 0.12\cite{lu_plasma-assisted_2014} &A$_{g}$\\

\hhline{|=|=|=|=|=|=|=|=|}
\multirow{2}{*}{Blue Phosphorene}
 & 610 & 550 & N/A & 1.00 & 1.00 & N/A &A$_1'$\\
\cline{2-8}
 & 439 & 439 & N/A & 0.92 & 0.43 & N/A &E$'$\\

\hhline{|=|=|=|=|=|=|=|=|}
\multirow{5}{*}{\thead{Armchair Graphitic\\Nanoribbon (N=6)}}
 & 1590 & 1555& 1595\cite{hans_agnr_2019} & 1.00 & 0.84& 0.33\cite{hans_agnr_2019}  & A$_1$\\
\cline{2-8}
 & 1352 & 1473& 1355\cite{hans_agnr_2019} & 0.02 & 0.08& 0.13\cite{hans_agnr_2019} & A$_1$\\
\cline{2-8}
 & 1317 & 1303& 1277\cite{hans_agnr_2019} & 0.18 & 0.50& 0.21\cite{hans_agnr_2019} & A$_1$\\
\cline{2-8}
 & 1237 & 1219& 1235\cite{hans_agnr_2019} & 0.58 & 1.00& 1.00\cite{hans_agnr_2019}  & A$_1$\\
\cline{2-8}
 & 457 & 539 &  451\cite{hans_agnr_2019}  & 0.03 & 0.04& 0.09\cite{hans_agnr_2019} & A$_1$\\
\hline
\end{tabular}
\caption{Comparison of frequencies ($\textrm{cm}^{-1}$) and normalized Raman intensities of experimentally observable vibrational modes between different computational approaches and experiments.\label{results}}
\end{table*}

\noindent where $H_{e-ph}=\frac{\partial V_{e-I}}{\partial Q_\nu}$ is the electron-phonon interaction (i.e. the derivative of the electron-ion potential with respect to phonon normal modes) and $\gamma_{e-ph}$ is the electron-phonon linewidth. Finally, the derivative of the polarizability can be obtained by substituting Eq. (\ref{eq2}) into Eq. (\ref{eq1}):
\begin{equation}
\frac{\partial \hat{\alpha}(\omega_l)}{\partial Q_\nu}=\sum\limits_{A,B}\frac{\langle i|\hat{H}_{e-p}|A\rangle\langle A|\hat{H}_{e-ph}| B\rangle\langle B|\hat{H}_{e-p}|f\rangle}{(\hbar\omega_l-\epsilon_A-i\gamma_{e-p})(\epsilon_A-\epsilon_B-i\gamma_{e-ph})}.
\label{freqplaczek}
\end{equation} 
For comparison one can also derive the expression of Raman scattering from time dependent perturbation theory as depicted in the Feynman diagram in Fig. \ref{compare}b:
\begin{equation}
K=\sum\limits_{A,B}\frac{\langle i|\hat{H}_{e-p}|A\rangle\langle A|\hat{H}_{e-ph}| B\rangle\langle B|\hat{H}_{e-p}|f\rangle}{(\hbar\omega_l-\epsilon_A-i\gamma_{e-p})(\hbar\omega_l-\epsilon_B-\hbar\omega_\nu-i\gamma_{e-ph})}.
\label{tdpert}
\end{equation} 
The main difference between Raman intensities in expression (\ref{freqplaczek}) and (\ref{tdpert}) manifests in their denominators. In expression (\ref{tdpert}) excitation energy dependence appears in both energy denominators. This means that by tuning the laser excitation energy, the Raman peak intensity will have two maxima, at $\epsilon_A$ and $\epsilon_B+\hbar\omega_\nu$, which are the incoming and outgoing resonance, respectively. The schematic representation of the incoming and outgoing resonances in the integrated Raman intensities can be seen in Fig. \ref{compare}c. 

In the expression (\ref{freqplaczek}) describing the frequency-dependent Placzek approximation only one of the denominators contains the excitation energy. This implies that only the incoming resonance will be found in the Raman spectra at $\epsilon_A$. Nevertheless, the amplitude of this resonance is approximately correct, because if the $\hbar\omega_l\approx\epsilon_A$ condition is satisfied, the second denominator can be written as $\hbar\omega_l-\epsilon_B$, thus expression (\ref{freqplaczek}) and (\ref{tdpert}) are approximately equivalent. The outgoing resonance, however, is not present in the approximate formula (\ref{freqplaczek}), therefore some differences can still be expected in the excitation profile as depicted in Fig. \ref{compare}c.

In practice, inclusion of the outgoing resonance in the second denominator is not possible within the frequency-dependent Placzek approximation, as the derivative is calculated after the sum over the virtual $|A\rangle$ states is performed in Eq. (\ref{eq1}). Therefore, the proper treatment of both energy denominators would require the calculation of both electron-photon and electron-phonon matrix elements, which is currently unavailable in most DFT codes. However, our method can be applied on top of just about any DFT software and is far more affordable than a full time dependent perturbation theory calculation of the same, while delivering very good accuracy for the incoming resonances as we show below.


The frequency-dependent polarizability tensor was evaluated using the built-in linear response algorithms\cite{gajdos_linear_2006} of VASP, within the local density approximation of density functional theory. Raman intensities were calculated with our own Python code\cite{RamPy}. Our code displaces atoms according to phonon normal modes symmetrically for both positive and negative directions and calls VASP to calculate the frequency-dependent dielectric tensor in the displaced geometries. After numerical differentiation our code computes the Raman spectra according to Eq. (\ref{raman}). Convergence test of this method and technical details of the calculations can be found in the Supplementary Material in sections S1 and S2.

To demonstrate the versatility of our method we considered several crystals with different lattice structures: %
face centered cubic (SiC), hexagonal close packed (ZnO), quartz (SiO$_2$), body centered cubic (anatase), two-dimensional (MoS$_2$, black and blue phosphorene) and quasi one-dimensional structures (armchair graphitic nanoribbon). We calculated vibrational frequencies and Raman intensities both with the Frequency-dependent \textit{ab initio} Placzek approximation (using VASP) and the static Placzek approximation (using QE) as shown in Table \ref{results}. Experimental data of frequencies and Raman intensities are also shown in Table \ref{results}. The irreducible representations corresponding to each normal mode are also noted in the last column. 

Comparing vibrational frequencies calculated by the two DFT codes, generally a good agreement can be found. Note, that since several vibrations are forbidden by symmetry in the Raman spectra, experimental values for the frequencies are limited to modes with measurable Raman intensity.

The overall good agreement of vibrational frequencies between theory and experiments does not propagate into the Raman intensities as presented in Table \ref{results}. Since absolute Raman intensities are usually difficult to compare, both theoretical and experimental peak intensities were normalized to the intensity of the highest peak. Comparing the results of the static Placzek approximation with experiments one can see that this approach mostly predicts which vibrational mode will have the highest intensity, but relative intensity ratios of smaller peaks are not accurate for most materials.

One possible explanation of the inaccuracy could be attributed to the fact that polarizations of incident and scattered light have a major effect on intensity ratios. To exclude this effect we took experimental data recorded on powdered samples containing multiple grains with various crystallographic orientation. In the case of the two-dimensional systems we compare to Raman spectra measured with unpolarized light. During the theoretical calculations, the Raman intensities were averaged over all directions for bulk crystals and parallel to the plane of the crystal for two-dimensional systems.

\begin{figure*}
\begin{center}
\includegraphics[scale=.4]{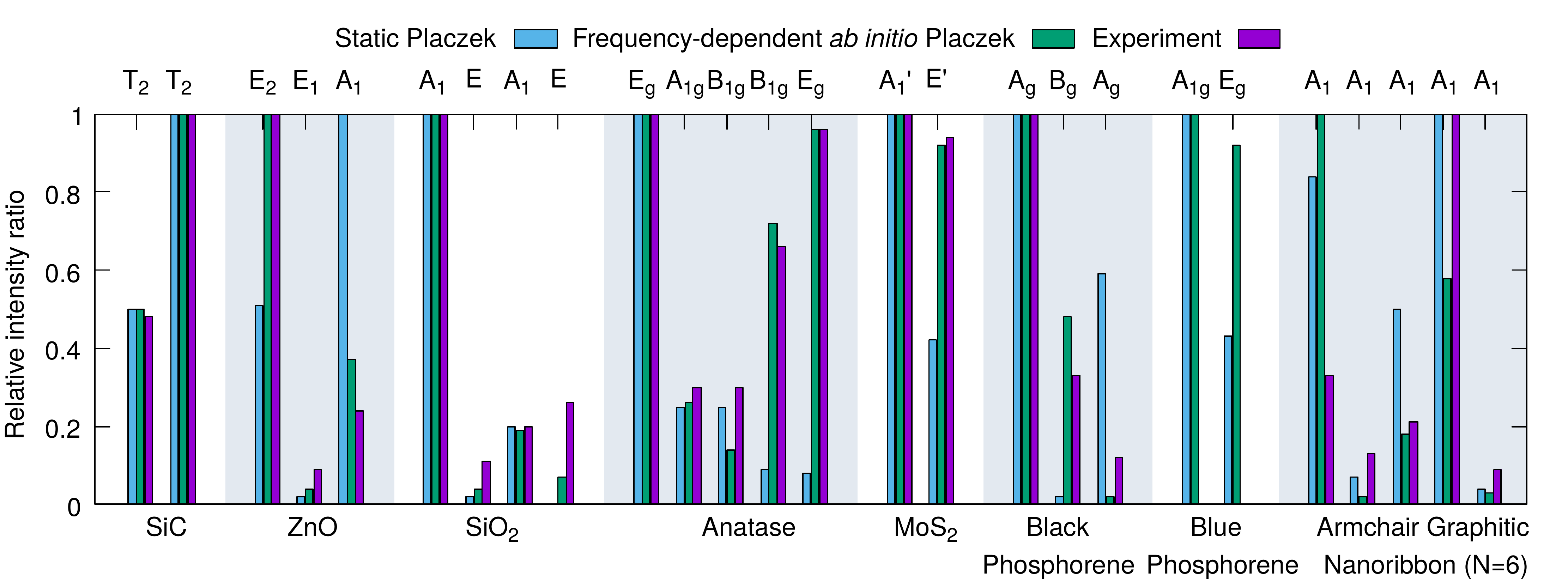}\\
\caption{Relative intenstiy ratio (compared to the peak with highest intensity) of peaks belonging to different materials and irreducible representations, results are equivalent of those shown in Table \ref{results}.\label{graph_results}}
\end{center}
\end{figure*}

Apart from polarization, Raman intensities are fundamentally dependent on the excitation energy, as different electron-hole pairs are excited at different laser energies. These electron-hole pairs are coupled to vibrations with various amplitudes, leading to different peak intensities. In off-resonance conditions (i.e. when the optical gap is large compared to the excitation energy), the Raman spectra can be modeled in the static limit. Measurable Raman signal can be detected due to Raman scattering through virtual electron-hole pairs with which resonances cannot occur, however, the absolute peak intensities are several orders of magnitude smaller compared to excitation energies where the resonance condition can be fulfilled. In the investigated systems shown in Table \ref{results} the gap is usually comparable to the energy of visible light, thus the difference in the intensity ratio between experiments and the static Placzek approximation can be attributed to the method neglecting the excitation energy dependence.

During our calculations using the Frequency-dependent \textit{ab initio} Placzek approximation, the polarizability can be obtained as a function of excitation energy within the whole visible spectrum. Using the previously described method, we calculated the Raman intensities as shown in Table \ref{results} and in Fig. \ref{graph_results}. Comparing these results with experimental values, very good agreement can be seen. Some small differences can still be found between the calculated and measured intensities, e.g. in the case of black phosphorene the intensity ratio of the two A$_g$ modes does not perfectly reproduce the experimentally observed values. These minor differences can be attributed to the inaccuracies of the applied \textit{ab initio} methods.

Treating the exchange and correlation energy using the local density approximation (LDA) always results in underestimated gap values, resulting in inaccurate resonance positions. A relatively simple method to correct this is the application of a scissor correction, that is, stretching the gap to the experimental value, but this would not necessarily alter the relative intensities of different vibrations. Alternatively, the electronic structure can be improved by taking into account many body corrections using the GW method.

 An additional source of inaccuracy is that the polarizability is calculated within the independent particle approximation (IPA), which excludes excitonic effects. The polarizability can be calculated more accurately using the Bethe-Salpeter equation (BSE)\cite{salpeter_relativistic_1951}, which is typically performed on top of a many-body GW calculation. In practice, however, the positions of electronic resonances are usually reproduced within margin of error by treating the polarizability at the LDA+IPA level, even for materials with large exciton binding energies.\cite{li_bulk_2012} This seemingly contradictory behavior is the result of the cancellation of two errors, as the difference between the LDA and GW quasiparticle gap usually matches the exciton binding energy. As a result, the peak energies in the optical absorption can be approximately reproduced on the LDA+IPA level, whilst the many-body corrections only change the amplitude of these peaks. In Raman spectroscopy the error introduced by the IPA is expected to be even less significant, as recent works show negligible difference between Raman spectra calculated on the LDA+IPA and GW+BSE level.\cite{gillet_ab_2017} Note, finally, that the theory behind the method we have presented in this work does not assume that the polarizability is calculated on the IPA level. Therefore, many-body effects can be included in our approach by replacing the IPA polarizability with the solution of the BSE. While this upgrade to our method presents an extremely high computational challenge, theoretical prediction of Raman intensities taking many-body effects into account should become feasible in the near future as high performance computing facilities improve.

In conclusion, we presented a hybrid classical-quantum model of resonance Raman spectroscopy by applying the frequency-dependent Placzek approximation to \textit{ab initio} quantum theory. We showed that this approach provides very good agreement with measurements for the relative intensities of the Raman modes. While the method is limited to describing incoming resonances, it is more affordable than a full time dependent perturbation theory calculation would be, and can be readily applied using any DFT code that has the built-in functionality of calculating the frequency-dependent polarizability matrix and the vibrational modes.

Support from the Hungarian National Research, Development and Innovation Office (NKFIH, Grant No. K-115608) is acknowledged. We acknowledge [NIIF] for awarding us access to resource based in Hungary at Debrecen. This research was supported by the National Research Development and Innovation Office of Hungary within the Quantum Technology National Excellence Program  (Project No. 2017-1.2.1-NKP-2017-00001). This work was completed in the ELTE Excellence Program (1783-3/2018/FEKUTSTRAT) supported by the Hungarian Ministry of Human Capacities. G. K. acknowledges support from the New National Excellence Program (UNKP) of the Ministry of Human Capacities in Hungary. V.Z. acknowledges support from the Graphene Flagship Project and the Computational Shared Facility at the University of Manchester. J. K. acknowledges the Bolyai and Bolyai+ program of the Hungarian Academy of Sciences.

\bibliography{bibliography}

\end{document}